\documentclass[epj]{svjour}
\usepackage{graphicx}
\begin{document}
\title{Experimental compaction of anisotropic granular media}
\author{Philippe Ribi\`ere, Patrick Richard, Daniel Bideau and Renaud Delannay}
\institute{Groupe Mati\`ere Condens\'ee et Mat\'eriaux, UMR CNRS 6626, Universit\'e de Rennes 1, Campus de Beaulieu, F-35042 Rennes cedex, France}
\date{}
\abstract{
We report on experiments to measure the temporal and spatial evolution of packing 
arrangements of anisotropic 
and weakly confined granular material, using high-resolution
$\gamma$-ray adsorption. In these experiments, the particle configurations start from an
initially disordered, low-packing-fraction state and under vertical solicitations evolve to a
dense state.
We find that the packing fraction evolution 
is slowed by the grain anisotropy
but, as for spherically shaped grains, can be  well fitted by a
stretched exponential. 
For a given type of grains, the characteristic times of relaxation and of convection are found to be of the same
order of magnitude.
On the contrary compaction mechanisms in the media
strongly depend on the grain anisotropy.
}
\authorrunning{P. Ribi\`ere et al}
\titlerunning{Experimental compaction of anisotropic granular media}
\maketitle
\section{Introduction}
Granular media are not thermal systems: the thermal
energy $k_{B}T$ is 
not involved in the evolution because
it is negligible compared to the variation of the
 gravitational energy during the motion of a grain. 
 Nevertheless they share with glasses  a great number of
 properties - such as off equilibrium dynamics, aging and hysteresis\ldots
It has even been suggested that the relaxation of
a granular medium under weak mechanical perturbations,
such as shaking, has a formal analogy with the slow dynamics of
out-of-equilibrium thermal systems~\cite{Edwards,Nicodemi}.
This analogy is based on the assumption that the most important
parameter in the system is the geometry and not the interaction
between particles or the driving energy.
Consequently, although mechanical agitation is neither stochastic 
nor isotropic, relaxation of granular media is often 
presented as an ideal system to study out-of-equilibrium
dynamics~\cite{Josserand2000}.

The first experiments of slow relaxation in granular media 
submitted to external mechanical excitations have been carried out in
Chicago~\cite{Chicago,Villarruel2000}. 
By implementing, 
at regular intervals, identical vertical taps of controlled 
intensity, they
show that the packing fraction of the media increases slowly
with the number of taps.
This has been done  for both 
spherical  ~\cite{Chicago} and anisotropic 
grains~\cite{Villarruel2000}.
For isotropic grains, 
 the evolution of the strongly confined media is correctly described 
by the  inverse of the logarithm of the number 
of taps described in equation~(\ref{eqn:Chicago}): 
\begin{equation}
\Phi=\Phi_{\infty}-(\Phi_{\infty}-\Phi_{0})\frac{1}{1+B\ln({1+t}/{\tau})}.
\label{eqn:Chicago}
\end{equation} 
 The four parameters adjusted during the
 fit are $\Phi_0$ the initial packing fraction, 
 $\Phi_\infty$ the final packing fraction (for $t\rightarrow\infty$),
 $\tau$, a characteristic time of the 
 evolution and  $B$ a parameter without any physical interpretation.

Other experiments have been carried out in quite different systems:
 compaction under shearing~\cite{Nicolas2000,Pouliquen2003} or under 
tapping 
but with a larger vessel~\cite{Philippe2002,PhilippePHD,Philippe2003,Richard2003}
- and thus a weaker confinement -
than the one used in~\cite{Chicago}.
In \cite{Philippe2002} the best fit found 
is not the one used by the Chicago group
but the so called KWW's fit
(Kohlrausch, Williams, Watts)~\cite{Kohlrausch1854,Williams1970}.
This law is a stretched exponential (Eq.~\ref{eqn:KWW}):
\begin{equation}
\Phi=\Phi_{\infty}-(\Phi_{\infty}-\Phi_{0}) \exp(-({t}/{\tau})^{\beta}),
\label{eqn:KWW}
\end{equation}
 where $ \tau$ is a characteristic time and
 $\beta$ measures the slowing down compared to
 the simple exponential law.
 Here $\Phi_0$ and $\Phi_\infty$ are not fit parameters but experimental
 data: $\Phi_0$ is the first point of the curve (the initial
 packing fraction) and $\Phi_\infty$ the mean value of the packing
 fraction at the stationary state.
 The tapping intensity is found to rule the characteristic time of
 relaxation according to an Arrhenius behavior  relation.
 All these characteristics are similar to those obtained
 in strong glassy systems and thus these results confirm
 the analogy between glasses and granular media explained 
 in the introduction.

The aim of this article is to check the validity 
of the results obtained in~\cite{Philippe2002,PhilippePHD}
with non-spherical shaped grains. 
Indeed, to reduce the number of
parameters, most previous studies deal with spherical
grains~\cite{Chicago,Nicolas2000,Pouliquen2003,Philippe2002,Philippe2003,PhilippePHD,Richard2003}.
Nevertheless actual granular material are far from being perfectly 
isotropic.
Grain anisotropy modify geometrical frustration and may
change packing behavior and convection
during compaction.

This paper is organized as follows. In the next section, the experimental
setup is described. Results on the relaxation and on the dynamics of compaction
are shown in section 3. 
The last section contains a summary of the findings
and conclusions.
\section{Experimental setup}
\begin{figure}[htb]
\begin{center}
\resizebox{0.75\columnwidth}{!}{
\includegraphics*{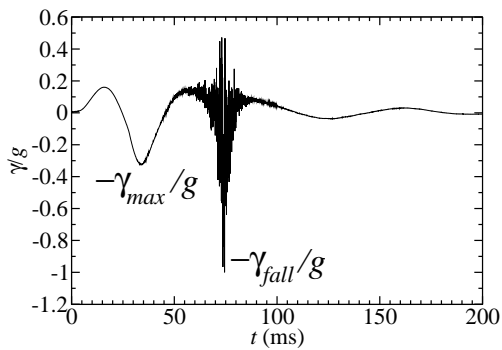}}
\caption{Example of dynamical signal monitored by the accelerometer
for $\Gamma\approx3.5$.
The acceleration created by the fall of the media on the bottom
$-\gamma_{fall}$ is different from the minimal acceleration applied
to the system $\gamma_{max}.$}
\label{fig:signal}
\end{center}
\end{figure}

\begin{figure}[htb]
\begin{center}
\resizebox{0.75\columnwidth}{!}{
\includegraphics*{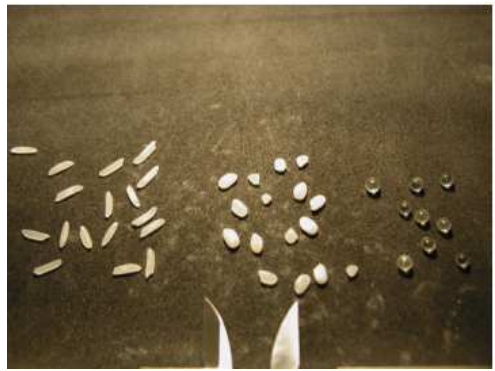}
}
\caption{Picture of the grains used: basmati rice (left), round rice (center)
and glass beads (right)}\label{fig:grains}
\end{center}

\end{figure}

The experimental setup consists of a glass cylinder of 
diameter $ D \approx 10\mbox{ cm}$ filled with about 600~g of grains
(corresponding to a height of roughly $10\mbox{ cm}$).
The method used to build our initial packing, already
used for spheres~\cite{Philippe2002,PhilippePHD}, is reproducible
(we obtain the same initial packing fraction and the same packing fraction evolution
for a given tapping amplitude).
{It consists of using two grids. The first one is placed 
at the bottom and inside  the cylinder. The grains are then poured
through the second one which is placed at the top the container.
Once the filling procedure is finished, the first grid is pulled
through the medium which is then dilated. This allows to obtain
a low initial packing fraction.}\\
The container is placed
on the plate connected to an electromagnetic exciter (LDS V406) which
induces a vertical displacement of the plate. The container is in this
way shaken at regular intervals ($\Delta t = 1 $s)
by vertical taps. Each tap is
created by selecting an entire period of
sine wave at a constant frequency $f= 30$~Hz.
The resulting motion of the whole system, monitored by an accelerometer
at the bottom of the container, is however more complicated than a
simple sine wave. At first, the system undergoes a positive acceleration
followed by a negative peak with a minimum equal to $-\gamma_{max}$ as showed
 in Figure~\ref{fig:signal}. Moreover, 
 it can be seen on this figure that the acceleration 
created by the fall 
of the media on the bottom of the cylinder, $-\gamma_{fall}$,
may be different from $-\gamma_{max}$.
Indeed we only used
taps for which the media took off from the bottom of the glass cylinder. It means 
the tap intensity is always chosen upon the lift-off threshold. 
After the
applied voltage stops, the system relaxes to its normal repose position.
Sequences of $10^4$ to $10^6$ taps are carried out with an intensity 
$\Gamma$ defined as follows: $\Gamma={\gamma_{max}}/{g}$ where $g$ is the 
gravitational acceleration. 
Here we call "time" the number of taps and the
"dynamics" is the succession of static equilibrium induced by the 
taps.\\
%
The average volume fraction in the bulk $\Phi$ is measured using the transmission ratio of 
$\gamma$-ray beam through the media. This setup allows to determine the vertical density 
profile too. The link between $\Phi$ and the ratio ${N}/{N_{0}}$, where $N$ and $N_{0}$
are the activities counted respectively by the detector with and without media, is:
$\Phi \approx \frac{1}{\mu}\ln({N}/{N_{0}}).$
In this equation, $\mu$ is the absorption coefficient of the beads determinated experimentally~\cite{Philippe2002,PhilippePHD}.
The collimated $\gamma$-beam is nearly cylindrical with a
diameter of $10$~mm and intercepts perpendicularly the vertical axis of
the cylindrical container. An acquisition-time of $60$~s was found  to be
a good compromise between the intrinsic uncertainty of a radioactive
beam and the total duration of an experiment. We then achieve a precision
$\Delta\Phi = \pm 0.003$. 
With the aim of limiting the duration of the experiments (from $20$ hours
to more than a month according to the total number of taps) and of
avoiding redundant information due to the very slow evolution of the
system, the measurements are spaced out in time, on a logarithmic scale,
with 2 measures of packing fraction profile and $50$ measures of 
mean packing fraction of the sample $\langle\Phi\rangle$
 per decade (except $10$
for the first decade).\\
The anisotropic grains used are rices of different shapes
(see Fig.~\ref{fig:grains}):
long grains (basmati rice) and short grains 
(round rice).
Results obtained with spheres in previous
works~\cite{Philippe2002,PhilippePHD} are also reported.
 In order to quantify the grain anisotropy each grain is assimilated
to an ellipsoid
and the mean ratio between the two axis is measured.
We obtain $2.5$ 
for the basmati rice and $1.5$  
for
the round rice. It is worth noting that the size distribution  is
larger for the round  rice (size dispersion around 
$20\%$) than for the basmati rice (less than $10\%$).

\section{Dynamics of compaction}
\subsection{influence of the shape on the initial and final packing fraction}

For disordered packings of spheres, computer simulations as well as experiments
have shown that the maximal packing fraction is set by the so-called random close packing
(RCP) ($\Phi\approx0.64$). Moreover the minimal packing fraction for a mechanically stable
sphere packing (the random loose packing) is often set to $\Phi\approx 0.55$.
Such estimations for disordered packings do not exist
for anisotropic-grain packings.
As mentioned above, 
the method used to build our initial packing is reproducible.
It gives
$\Phi\approx 0.56$  for round rice and
$\Phi\approx 0.55$  for basmati rice, which means
that, in our experiments,  the more anisotropic the grains are,
the less important the initial packing fraction is.
This can be explained by the fact that
arches - and thus large pores - form more easily
 with anisotropic grains.
\begin{figure*}[hbt]
\begin{center}
\includegraphics*[width=15cm]{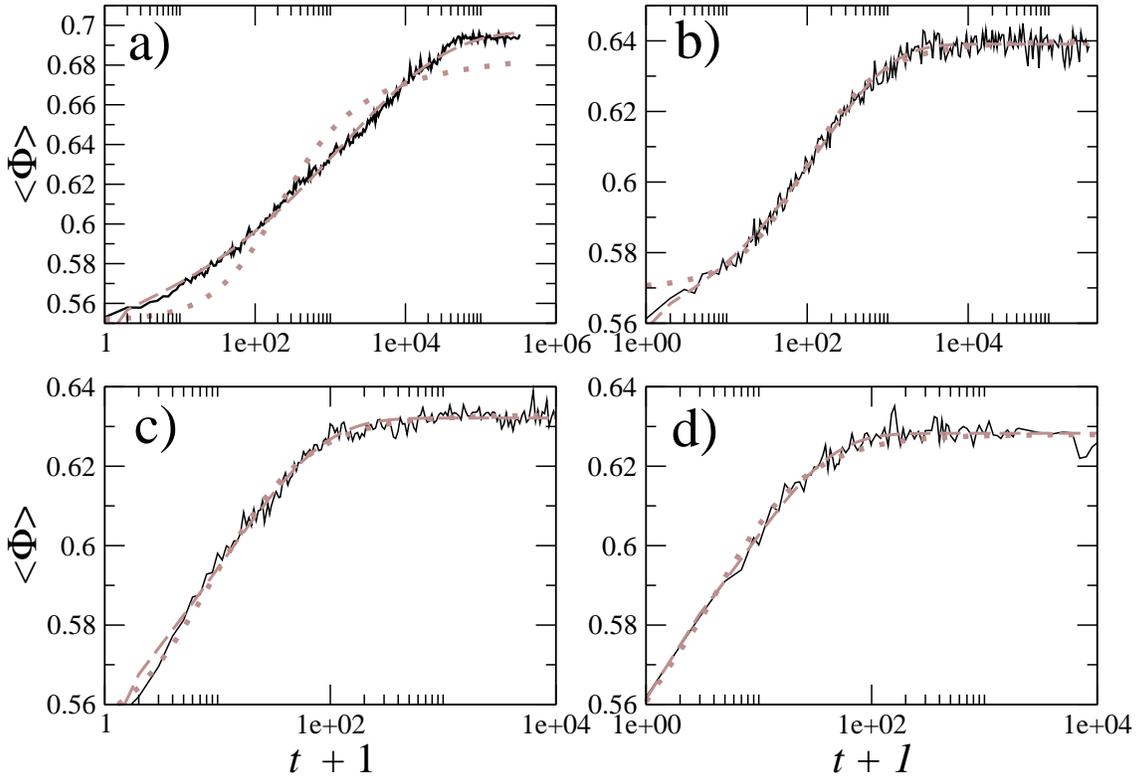}
\caption{Evolution of the packing fraction of
a) basmati rice with $\Gamma=2.4$, b)
round rice with $\Gamma=2.4$, c)
basmati rice with $\Gamma=6$ and d)
round rice with $\Gamma=6$.
Chicago group's fit (dotted line) and KWW's fit (dashed line) are 
reported for each curve. {For all the fits we used the Levenberg-Marquardt algorithm
and the parameters $\Phi_0$ and $\Phi_\infty$ are set to the 
experimental values}.}
\label{fig:compalog}
\end{center}
\end{figure*}
%
%
Figure~\ref{fig:compalog} reports the evolution of the packing fraction with the number of taps
for the two kinds of rice and for $\Gamma=6$  and $2.4$.
The first observations is that, whereas final packing fractions
for round rice are comparable to what is obtained for 
compaction of sphere packings, they can be much higher for
the basmati rice.
Nevertheless we do not observe,
contrary to Villarruel et al~\cite{Villarruel2000},
an ordered phase in which the grains align vertically .
 This order seems to be due to the strong lateral confinement. 
Indeed in~\cite{Villarruel2000} the length of the rods used is the same as the diameter 
of the cylinder so the steric constraints are huge. In our experiments
the smaller ratio
 between the diameter of the cylinder
 and the length of the grain is about $25$.
 Nevertheless, as explained below, we can observe for given values of $\Gamma$, 
 an order which is different from that observed in~\cite{Villarruel2000}.
\subsection{Dynamics of compaction for basmati rice}
\begin{figure}
\begin{center}
\resizebox{0.75\columnwidth}{!}{%
\includegraphics*{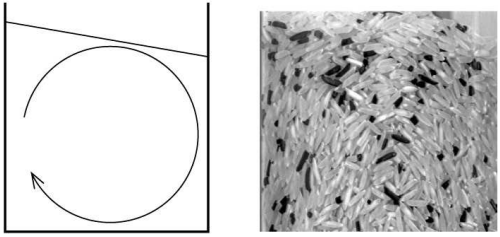}
}
\caption{sketch of the convection obtained for basmati rice and for $\Gamma=6$ (left)
and snapshot of the media during relaxation (right) where convection roll can be 
clearly seen on the side walls}\label{fig:vortex1}
\end{center}
\end{figure}

Depending on the value of $\Gamma$, two types of phenomena can be
observed in basmati rice: 
 for high values of tapping intensity
$\Gamma$, convection is observed 
in the whole media whereas for low values of $\Gamma$ convection
is localized in a part of the packing and is not steady.
As it will be shown, compaction mechanisms strongly depend
on the
convection and thus on $\Gamma$.
It should be pointed out that we only consider
values of $\Gamma$ larger than the 
lift-off threshold~\cite{Philippe2003}.\\
For high tapping intensity (typically $\Gamma > 3$) convection
takes place
within the whole medium. After about ten taps, two convection rolls
appear but this situation becomes unstable.
One of the rolls progressively disappears and after this transient the
whole medium is fulfilled by only one roll and
the free surface of the medium is tilted from the horizontal
(for example about $20^\circ$ for $\Gamma=6$).
This can be observed
at the vessel wall as showed in Figure~\ref{fig:vortex1}. 
With regular acquisition of images 
after each mechanical perturbation, a characteristic time for
the evolution of convection rolls, $\tau_{conv}$, can be extracted. 
Indeed, 
we can track grain movements at the vessel sides and measure how long
it takes a grain to revolve around the
center of the convection roll. 
For basmati rice submitted to taps of intensity 
$\Gamma=6$ we find $\tau_{conv} \approx 50$.
The evolution of
packing fraction during this experiment, 
is reported  in Figure~\ref{fig:compalog}c.
Chicago group's fit and KWW's fit 
{(using the Levenberg-Marquardt algorithm)}
both correctly describe the experimental evolution
and 
{they give extremely similar 
correlation coefficients ($\approx 0.991$)}.
However, the evolution of the former fit near the stationary state 
is slower than the
experimental result.
The fit  
provides the following values: $\tau=1150$ and $B=165$
for Chicago group's law 
and $\tau=30 $ and $ \beta=0.58 $ for the KWW's law.
The value of $\tau_{KWW}$ is of the same order of magnitude
than $\tau_{conv}$ and this remains valid for 
all our experiments done for $\Gamma > 3$.
Note that for the fits the parameters
$\Phi_\infty$ and $\Phi_0$ are set to the experimental values.

\begin{figure}
\begin{center}
\resizebox{0.75\columnwidth}{!}{%
\includegraphics*{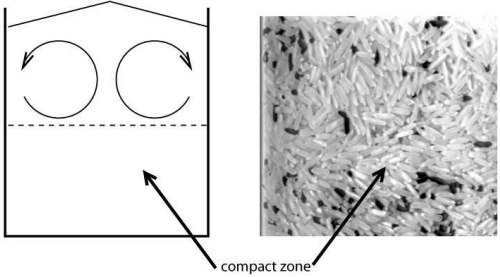}
}
\caption{sketch of the convection obtained for basmati rice and for $\Gamma=2.4$ (left)
and snapshot of the media during relaxation (right) where the compact and ordered
part can be
clearly seen on the side walls}\label{fig:vortex2}
\end{center}
\end{figure}

The behavior of our system is totally different for taps of
low intensity ($\Gamma<3$).
Two convection rolls 
roughly equally-sized  and
localized near the free surface 
appear.
The other part of this packing (below
the rolls) is not submitted to convection.
As for the high tapping intensity case, this
situation is unstable. Contrary to the previous
case the two rolls do not merge: their size
continuously decreases
until they disappear. 
For basmati rice 
and for a solicitation of intensity $\Gamma=2.4$, the convection
is stopped in most of the packing after $\tau\simeq1500$.
Convection rolls do not exist anymore, except near the air-media interface (see Fig.~\ref{fig:vortex2}).
Another interesting point is that  an ordering process is observed at the bottom of the
vessel: the convection rolls compact the lowest part of the packing
and align the rice grains horizontally. 
{This explains why the packing fractions obtained
can be so high}.
This order creation 
is different from that observed by Villarruel et al~\cite{Villarruel2000}
where the grains are preferentially oriented vertically.
Thus the order observed in~\cite{Villarruel2000} is created
by the side walls whereas the order observed in the present work
is created by convection. Note that the aspect ratio of the grains
is probably also an important parameter.
 The evolution of packing fraction during this experiment is showed 
 Figure~\ref{fig:compalog}a.
 On this figure, it can be  seen that the fit from Chicago group's law (logarithmic law) does
 not describe
 the experimental evolution very well.
 Indeed, Chicago group's law increases too fast 
 before the steady state and reaches it too late.
 The parameters of Chicago group's law are $\tau=375$ and $B=2$
 ({with a correlation coefficient of 0.972}).
 {We recall that for the fit we set the values of 
 $\Phi_\infty$ 
 equal to the experimental final packing fraction.
 A better correlation coefficient 
 can be obtained (0.996) if $\Phi_\infty$ is a free parameter
 but its value  
 (0.86) is then unrealistic and well above the value of 
 the plateau observed in Figure~\ref{fig:compalog}a.}
 So for this law, the parameter $\tau$ for $\Gamma=2.4$ is smaller than $\tau$ for $\Gamma=6$ whereas the evolution
 slows down with the decrease of $\Gamma$. 
This characteristic time which comes from Chicago group's fit does not show a
physical behavior and does not describe our results. 
On the contrary 
KWW's law correctly fits the experimental evolution with the following parameters:
$\tau=1380$ and $\beta=0.31$ ({with a correlation coefficient of 
0.998}).
Like for the high-intensity solicitations, $\tau_{KWW}$
the time
 extracted from KWW's fit and the characteristic time
 of the convection rolls are similar. So in all our experiment,
$\tau_{KWW}$ can be correlated with a real time of evolution.\\
It should be pointed out that such kind of correlation
is not possible in the experiments of Chicago because 
the use of a very narrow vessel prevents convection.
Therefore the convection mechanisms are not the same in the
two setup.
\subsection{dynamics of compaction for round rice}
For round rice, we did not observe the same change
of behavior for $\Gamma \approx
3$. The convection always took place in the media, as for spherical
beads~\cite{Philippe2002,Philippe2003}. And, as for spherical beads, KWW's fit describes better
the experimental evolution of the packing fraction than Chicago group's law
(see Figures~\ref{fig:compalog}b and \ref{fig:compalog}c).
 However we could see a strong decrease of the convection speed
for intensity taps under $\Gamma \approx 3$. More results about convection 
can be found in~\cite{Ribiere2004}.
\subsection{Dependence of the relaxation on $\Gamma$}
\begin{figure}[htb]
\begin{center}
\resizebox{0.75\columnwidth}{!}{%
\includegraphics*{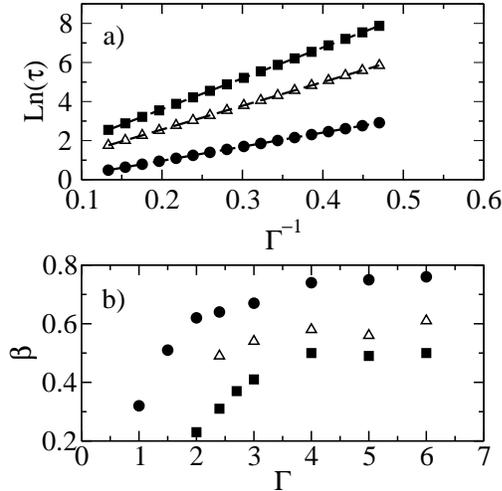}
}
\caption{
(a) Characteristic time of the stretched exponential as a function
of the inverse of the tapping intensity $\Gamma$
for glass beads (filled circles), round rice (open triangles)
and basmati rice (filled squares).
(b) $\beta$  exponent of the stretched exponential fits
as function of the tapping intensity $\Gamma$ for 
glass beads (filled circles), round rice (triangles)
and basmati rice (filled squares).
Glass bead data are taken from~\cite{Philippe2002} and \cite{PhilippePHD}.}
\label{fig:tau_beta}
\end{center}
\end{figure}

For our experimental setup, KWW's law
is always the best fit. 
For all the grains used the parameter $\tau$ 
follows an 
Arrhenius 
 behavior 
(see Fig.~\ref{fig:tau_beta}a) 
\begin{equation}
\tau(\Gamma)=\tau_0\exp\left(\frac{\Gamma_0}{\Gamma}\right).
\end{equation}
The values of the parameters are
$\tau_{0} \approx  1.06$  and  $\Gamma_{0} \approx 7$ for glass beads, 
$1.14$ and $12$ for short rice and $1.56$ and $16$ for long rice. Even if the interpretation
of $\tau_{0}$ is not easy, the value of $\Gamma_{0}$ is linked to the height of energy
step in the phases space. Indeed, more anisotropic are the beads, larger is the
distribution of energy in phases space and so greater must be the typical intensity of
a tap to explore this space. 

The evolution of $\beta$ is reported in Figure~\ref{fig:tau_beta}b.
It is qualitatively the same for 
the different kinds of grains. For a given value of tapping intensity 
$\Gamma$, the difference to a perfect exponential ($\beta=1$)
is more important for anisotropic grains. The stretching
of the exponential, and thus the coefficient is $\beta$,
is due to a wide characteristic time dispersion. Since the 
range of metastable states is wider for anisotropic particles
the characteristic times are more dispersed and thus
the coefficient $\beta$ lower.

\begin{figure}[htb]
\begin{center}
\resizebox{0.75\columnwidth}{!}{%
\includegraphics*{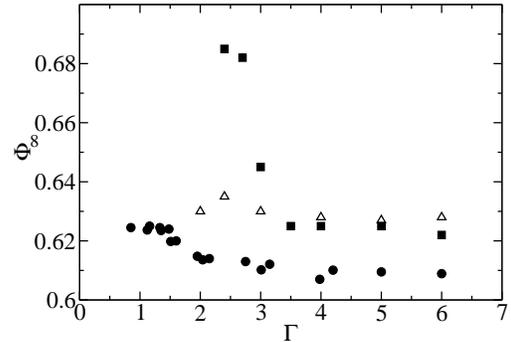}
}
\caption{Final packing fraction as a function of the tapping intensity
$\Gamma$ for glass beads (filled circles), round rice (open triangles)
and basmati rice (filled squares).
Glass bead data are taken from~\cite{Philippe2002} and \cite{PhilippePHD}.
}\label{fig:compafinale}
\end{center}
\end{figure}

The last quantity reported is the final packing fraction (Fig.~\ref{fig:compafinale}).
Although the general shape of the curves 
is similar, the quantitative measures depend on grain shape. 
Moreover the change of behavior observed for glass
beads~\cite{Philippe2002,Philippe2003} around
$\Gamma\approx2$ can also be seen
for basmati and round rice. This change is 
very important for basmati rice at $\Gamma\approx3$.
The study of a perfect non-inelastic solid on a plate with a sinus motion shows that the flying time does not depend of the solid
mass~\cite{Mehta1990}. This explain that the behavior of the glass media and the rice is qualitatively the same with $\Gamma$. Nevertheless,
a small difference of the threshold between the two regimes is observed  but it is due to the change in friction coefficient
between the grains and the vessel. 

\section{Conclusion}
 We have studied the effect of grain anisotropy on granular compaction
under vertical tapping, and weak lateral confinement.
 Using a $\gamma$-ray adsorption apparatus, we have reported the evolution
of the packing fraction as a function of the number of taps.
We observe that the main features of granular compaction do not
qualitatively depend on grain shape. In particular, the packing fraction
evolution can be well fitted by a stretched exponential in all the cases.
The fit parameters as well as the final packing fraction have the same
qualitative behavior and
the fit parameter $\tau$ is found to be correlated to the
experimental characteristic time of the convection.
Nevertheless we observed that the convection in the granular media,
and thus the compaction mechanisms,
strongly
depend on grain anisotropy.
 

\end{document}